\documentclass[12pt]{iopart}
\usepackage{iopams}
\usepackage{graphicx}
\input amssym.def \input amssym

\begin{document}

\title[]{Balanced Tripartite Entanglement, the Alternating Group $A_4$ and the Lie Algebra $sl(3,\mathbb{C})\oplus u(1)$ }

\author{Michel Planat$^{1,4}$, Peter Levay$^{2,4}$ and Metod Saniga$^{3,4}$}

\address{$^1$ Institut FEMTO-ST, CNRS, 32 Avenue de
l'Observatoire,\\ F-25044 Besan\c con, France. }
\vspace*{.1cm}

\address{$^2$ Departement of Theoretical Physics, Institute of Physics, \\Budapest University of Technology and Economics, H-1521 Budapest, Hungary}
\vspace*{.1cm}

\address{$^3$ Astronomical Institute, Slovak Academy of Sciences,\\ SK-05960 Tatransk\`{a} Lomnica, Slovak Republic.}
\vspace*{.1cm}

\address{$^4$ Center for Interdisciplinary Research (ZiF), University of Bielefeld,\\
D-33615 Bielefeld, Germany}

\begin{abstract}
We discuss three important classes of three-qubit entangled states and their
encoding into quantum gates, finite groups and Lie algebras. States of the GHZ
and W-type correspond to pure tripartite and bipartite entanglement,
respectively. We introduce another generic class B of three-qubit states, that
have balanced entanglement over two and three parties. We show how to realize
the largest cristallographic group $W(E_8)$ in terms of three-qubit gates (with
real entries) encoding states of type GHZ or W [M. Planat, {\it Clifford group
dipoles and the enactment of Weyl/Coxeter group $W(E_8)$ by entangling gates},
Preprint 0904.3691 (quant-ph)]. Then, we describe a peculiar ``condensation" of
$W(E_8)$ into the four-letter alternating group $A_4$, obtained from a chain of
maximal subgroups. Group $A_4$ is realized from two B-type generators and found
to correspond to the Lie algebra $sl(3,\mathbb{C})\oplus u(1)$. Possible
applications of our findings to particle physics and the structure of genetic
code are also mentioned.

\end{abstract}


\noindent

\section{Introduction}

Tripartite aggregates and interactions frequently occur in the natural world.
As a first example, it is well known that ordinary matter consists of atoms
whose nuclei are made of protons and neutrons, which are themselves made of the
lighest quarks $u$ and $d$. A proton consists of a triplet $uud$ and a neutron
consists of a triplet $ddu$. Thus, our present universe is made of three types
of stable particles, of spin $\frac{1}{2}$, i.e. electrons $e$ and $u$ and $d$
quarks. According to the standard model, there also exists four heavier quarks
(among them the strange spin $\frac{1}{2}$ quark $s$), that combine to form
unstable composite particles called hadrons, in quark-antiquark pairs (mesons)
or three-quark states (baryons). Mathematically, these composite particles are
described using the representations of the Lie algebra $su(3)$, in a model
named the eightfold way by Gell-Mann and Ne'eman \cite{Carruthers1966}. An old
instance goes back to the beginning of chemistry. Among the numerous precursors
of Mendeleev, D\"{o}bereiner was the first to classify chemical elements into
triads \cite{Kibler07}.

A second relevant example is the genetic code (or amino acid code), that refers
to the system of passing from DNA and RNA into the synthesis of proteins. It
was discovered in 1961 by Crick {\it et al.} that the genetic code is a triplet
code, made of elementary units of information called codons. There are $64$
codons made of four building block bases $A$, $U$, $G$ and $C$ that encode $20$
aminoacids. A chain of subalgebras of the Lie algebra $sp(6)$ was proposed for
explaining the high degeneracy of the code \cite{Hornos93}. See also the
modeling of the genetic code based on quantum groups in \cite{Sorba98} and
related papers.

Our third example is quantum information theory. The term {\it black hole
analogy} has been coined for featuring the relationship between some stringy
black hole solutions and three-qubit states \cite{Duff07,Levay08}. Presumably,
this analogy stems from the structure of the largest cristallographic group
$W(E_8)$, of cardinality $696~729~600$, which one of the authors succeeded in
representing in terms of several three-qubit gates \cite{Planat09}.


Among the various forms of three-qubit entanglement, a first classification
based on SLOCC (stochastic local operations and classical communications) leads
to entangled states of the type GHZ and W. The former possess pure (and
maximal) three-qubit entanglement and any tracing out about one party destroys
all the entanglement. The latter possess equally distributed (and maximal)
bipartite entanglement, but no tripartite entanglement. A finer classification
is based on local unitary equivalence \cite{Acin00}. In this paper, we are
especially interested in a class of entangled three-qubit states displaying
equally distributed entanglement about three and two parties. Such states were
already encountered in the context of CPT symmetry \cite{PlanatCPT09}. Here,
they occur when one ``condenses" the three-qubit representation of $W(E_8)$ to
the alternating group $A_4$, through an appropriate chain of maximal subgroups.
The Lie subalgebra of rationals obtained from the generators of $A_4$ is found
to be $sl(3,\mathbb{C}) \oplus u(1)$. Going upstream in the group sequence, one
arrives at a representation of the symmetric group $S_4$, with attached Lie
algebra $sl(3,\mathbb{C})\oplus sl(2,\mathbb{C})\oplus u(1)\oplus u(1)$, that
may play a role in the understanding of elementary particles \cite{Ma87, Ma06}.


In this paper, we expose our new findings about $B$-type entanglement
(Sec.\,2), the generation of $W(E_8)$ with entangling matrices and the
embedding of specific permutation groups $S_4$ and $A_4$ (Sec.\,3). A novel
three-qubit realization of $sl(3,\mathbb{C}) \oplus u(1)$ and its
generalization is described in Sec.\,4. A rudimentary explanation of Lie groups
and algebras is given in the appendix.

Many calculations are performed by using the abstract algebra software Magma
\cite{MAGMA}. A few papers relating Lie algebras and quantum information theory
have already been published \cite{Bandyo02}-\cite{Domotor09}.

\section{$B$-type three-qubit quantum entanglement}

One efficient measure of two-qubit entanglement is the tangle $\tau=C^2$, where
the concurrence reads $C(\psi)=|\langle \psi |\tilde{\psi}\rangle|$. The
flipped transformation $\tilde{\psi}=\sigma_y \left|\psi^{\ast}\right\rangle$
applies to each individual qubit and the spin-flipped density matrix
$\tilde{\rho}=(\sigma_y \otimes \sigma_y)\rho^{\ast}(\sigma_y \otimes
\sigma_y)$ follows \cite{Coffman00}. Explicitly,
%
$$C(\rho)=\mbox{max}\left\{0,\sqrt{\lambda_1}-\sqrt{\lambda_2}-\sqrt{\lambda_3}-\sqrt{\lambda_4}\right\}, $$
%
where the $\lambda_i$ are (non-negative) eigenvalues  of the product $\rho \tilde{\rho}$, ordered in decreasing order.

Roughly speaking, two pure multiparticle quantum states may be considered as
equivalent if both of them can be obtained from the other by means of
stochastic local operations and classical communication (the SLOCC group)
\cite{Dur00}. There are essentially two inequivalent classes of three-qubit
entangled states, with representative $\left|GHZ
\right\rangle=\frac{1}{\sqrt{2}}(\left|000 \right\rangle+\left|111
\right\rangle)$ (for the GHZ class) and $\left|W
\right\rangle=\frac{1}{\sqrt{3}}(\left|001 \right\rangle+\left|010
\right\rangle+\left|100 \right\rangle)$ (for the $W$-class). For {\it
measuring} the entanglement of a triple of quantum systems $A$, $B$ and $C$,
one may calculate the amount of {\it true} three-qubit entanglement from the
SLOCC invariant three-tangle \cite{Wootters00}
$$\tau^{(3)}=4\left|d_1-2d_2+4d_3\right|,$$
$$ d_1=\psi_{000}^2\psi_{111}^2+\psi_{001}^2\psi_{110}^2+\psi_{010}^2\psi_{101}^2+\psi_{100}^2\psi_{011}^2,$$
$$d_2=\psi_{000}\psi_{111}(\psi_{011}\psi_{100}+\psi_{101}\psi_{010}+\psi_{110}\psi_{001})$$
$$+\psi_{011}\psi_{100}(\psi_{101}\psi_{010}+\psi_{110}\psi_{001})+\psi_{101}\psi_{010}\psi_{110}\psi_{001},$$
$$d_3=\psi_{000}\psi_{110}\psi_{101}\psi_{011}+\psi_{111}\psi_{001}\psi_{010}\psi_{100}, $$
%
as well as the amount of two-qubit entanglement between two parties, by tracing out over partial subsystems $AB$, $BC$ and $AC$.

For a two-qubit state $\left| \psi\ \right\rangle=\alpha \left|00\
\right\rangle + \beta\left|01\ \right\rangle+ \gamma\left|10\ \right\rangle+
\delta\left|11\ \right\rangle$, the concurrence is
$C=2\left|\alpha\delta-\beta\gamma\right|$, and thus satisfies the relation
$0\le C \le 1$, with $C=0$ for a separable state and $C=1$ for a maximally
entangled state.

The three-qubit entangled state $\left|GHZ \right\rangle$ is maximally
entangled, with three-tangle $\tau^{(3)}=1$ and all two-tangles vanishing; that
is, whenever one of the qubits is traced out, the remaining two are completely
unentangled. On the other hand, the entangled state $\left|W \right\rangle$ has
$\tau^{(3)}=0$, but it maximally retains bipartite entanglement \cite{Dur00}.

Refinements on the above classification may be obtained if one classifies the
three-qubit state up to local unitary equivalence (the LU group) \cite{Acin00}.
Thus, if one singles out the first party $A$, a generic state of three qubits
depends, up to LU, on five parameters
$$\left|\psi \right\rangle=\lambda_0 \left|000 \right\rangle +\lambda_1 e^{i \phi}\left|100 \right\rangle+\lambda_2 \left|101 \right\rangle+\lambda_3 \left|110 \right\rangle+\lambda_4 \left|111 \right\rangle,$$

$$\lambda_i>0,~\sum_{j=0}^4 \lambda_i^2=1 ~\mbox{and}~0 \le \phi \le \pi.$$
In the sequel, we are interested in entangled states of the $B$-class, where
$\lambda_1=0$, with a representative
\begin{equation}
$$\left|B \right\rangle=\frac{1}{2}( \left|000 \right\rangle+\left|101 \right\rangle+ \left|110 \right\rangle+ \left|111 \right\rangle).$$
\label{balanced}
\end{equation}
The three-tangle of the $B$-state is $\tau^{(3)}=\frac{1}{4}$ and the density matrices of the bipartite subsystems are
\scriptsize
$$\rho_{BC}=\frac{1}{4}\left(\begin{array}{cccc} 1 & 0 & 0 & 0 \\0 & 1 & 1 & 1 \\ 0 & 1 & 1 & 1 \\0 & 1 & 1 & 1\\ \end{array}\right),~
\rho_{AB}=\frac{1}{4}\left(\begin{array}{cccc} 1 & 0 & 0 & 1 \\0 &0 & 0 & 0 \\ 0 & 0 & 1 & 1 \\1 & 0 & 1 & 2\\ \end{array}\right),~
\rho_{AC}=\frac{1}{4}\left(\begin{array}{cccc} 1 & 0 & 0 & 1 \\0 & 0 & 0 & 0 \\ 0 & 0 & 1 & 1 \\1 & 0 & 1 & 2\\ \end{array}\right).~$$
%
\normalsize

The set of eigenvalues $\left\{\frac{1}{16}(3+2\sqrt{2}),
\frac{1}{16}(3-2\sqrt{2}),0,0\right\}$ is uniform over the subsystems with
two-tangles $\tau_{AB}=\tau_{AC}=\tau_{BC}=\frac{1}{4}$. Similarly, the linear
entropies $\tau_{A(BC)}=\tau_{B(AC)}=\tau_{C(AB)}=\frac{3}{4}$ are the same
(see \cite{Coffman00} for the meaning of linear entropies such as
$\tau_{A(BC)}= \tau^{(3)}+\tau_{AB}+\tau_{AC}$). Thus, the entanglement measure
for two parties equals the entanglement measure for three parties. This equal
balance of the entanglement for two or three parties justifies our notation for
the $B$-class \footnote{The $B$-states are denoted CPT states in our previous
work \cite{PlanatCPT09}. Choudhary and coworkers \cite{Rahaman09} computed the
local realistic violation of the inequality (given in Eq. (3) of their paper)
for the generic state $\left|B \right\rangle$ and found the value $0.608723$, a
big violation compared to $0.175459$ for the generic GHZ state and $0.192608$
for the generic $W$ state. }.

\section{Three-qubit entanglement and the crystallogaphic group $W(E_8)$}
\label{Entangl3}

Recently, by studying the Clifford group on two and three qubits, we discovered
several eight-dimensional orthogonal realizations of the largest
crystallographic group $W(E_8)$, and of its relevant subgroups. As described in
papers \cite{Planat09,PlanatCPT09}, these representations find their kernel in
two-qubit entanglement and the following orthogonal matrix
\begin{equation}
S_2=\frac{1}{2}\left(\begin{array}{cccc} 1 & -1 & 1 & 1 \\1 & 1        & -1 & 1 \\1 & -1 & -1 & -1 \\1 & 1 & 1 & -1\\ \end{array}\right),~~
\left(\begin{array}{ccc} + & - & - \\- &+        & -  \\ - & - & +  \\+ & + & + \\ \end{array}\right),
\label{gateS2}
\end{equation}
that encodes the joint eigenstates of the triple of observables.
\begin{equation}
\left\{\sigma_x\otimes \sigma_z,\sigma_z\otimes \sigma_x,\sigma_y\otimes \sigma_y\right\}.
\label{2triple}
\end{equation}
Rows of the second matrix contain the sign of eigenvalues $\pm 1$ of the triple
of observables, and a row of the first matrix corresponds to a joint eigenstate
[e.\,g. the first row corresponds to the state
$\frac{1}{2}(\left|00\right\rangle-\left|01\right\rangle+\left|10\right\rangle+\left|11\right\rangle$
with eigenvalues $(1,-1-1)]$.

To abound in this claim, let us consider the following triple of three-qubit observables
\begin{equation}
\sigma_z \otimes\left\{\sigma_x\otimes \sigma_z,\sigma_z\otimes \sigma_x,\sigma_y\otimes \sigma_y\right\},
\label{3triple}
\end{equation}
that follows from (\ref{2triple}) by adjoining the tensor product $\sigma_z$ at
the left hand side. Eigenstates of (\ref{3triple}) may be used for encoding the
rows of the following orthogonal matrix
\footnotesize
\begin{equation}
S_3=\frac{1}{2}\left(\begin{array}{cccccccc} 0&0&0&0 &1&1&1&-1 \\1&1&1&-1 &0&0&0&0\\0&0&0&0 &1&1&-1&1\\1&-1&1&1 &0&0&0&0\\
1&1&-1&1 &0&0&0&0 \\-1&1&1&1 &0&0&0&0\\0&0&0&0 &1&-1&1&1\\0&0&0&0 &-1&1&1&1\\ \end{array}\right),
\end{equation}
\label{gateS3}
\normalsize
and to generate the derived subgroup $W'(E_8)\cong O^+(8,2)$ of order
$348~364~800$ [recall that $O^+(8,2)$ is the general eight-dimensional
orthogonal group over $GF(2)$]
\begin{equation}
W'(E_8)\cong \left\langle \sigma_x \otimes S_2,S_3\right\rangle.
\label{E8}
\end{equation}
Replacing the $S_3$ state by the GHZ-type generator $b$ whose explicit form is
given by Eq. (18) of \cite{Planat09}, one gets $W'(E_7)\cong \left\langle
\sigma_x \otimes S_2,b\right\rangle$. Indeed many important subgroups of
$W(E_8)$ may be realized by means of the appropriate orthogonal generators.

Here one focuses on a sequence of subgroups leading to a specific
representation of the four-letter alternating group $A_4$ (as well as the
symmetric group $S_4$) and a representation of the Lie algebra
$sl(3,\mathbb{C})$ (as well as its more general parent). The relevant sequence
is
\begin{equation}
W'(E_8)\supset W'(E_7)\supset W'(E_6)\supset G_{648}\supset S_4\supset A_4.
\label{sequ}
\end{equation}
Starting from $W'(E_8)$ [as in (\ref{E8})], one looks at the maximal subgroups.
One of the three subgroups of the largest cardinality is isomorphic to
$W(E_7)$, of order $2~903~040$ \footnote{The second largest subgroup of
$W(E_8)$ is the real Clifford group $\mathcal{C}_3^+$, of order $2~580~480$
studied in \cite{Planat09,PlanatCPT09}.}. Then, in the derived subgroup
$W'(E_7)$, one takes the largest maximal subgroup $W(E_6)$, of order $51~840$.
Among the five maximal subgroups of $W'(E_6)$, two of them have the cardinality
$648$; one selects the one isomorphic to the semi-direct product $G_{648}=Z_2^7
\rtimes S_4$ \footnote{The maximal subgroup of the largest cardinality in
$W'(E_6)$ is isomorphic to the perfect group $M_{20}=Z_2^4 \rtimes A_5$ of
order $960$, and is described in \cite{PlanatJorrand08, PlanatSole08}.}.
Finally, one is interested in the subgroup $S_4$ of $G_{648}$, as well as in
its derived subgroup $A_4$.

%
%
The alternating group $A_4$ may be realized by means two orthogonal generators
$x_{A_4}$ and $y_{A_4}$, whose rows are similar up to a permutation, and encode
three-qubit states of the $B$-type, with similar two- and three-tangles as it
results from straightforward calculations.

\footnotesize
\begin{eqnarray}
&x_{A_4}=\frac{1}{2}\left(\begin{array}{cccccccc} 0&1&-1&-1 &0&0&1&0 \\0&1&1&-1 &0&0&-1&0\\0&1&1&1 &0&0&1&0\\-1&0&0&0 &1&1&0&-1\\
-1&0&0&0 &1&-1&0&1 \\-1&0&0&0 &-1&1&0&1\\-1&0&0&0 &-1&-1&0&-1\\0&1&-1&1 &0&0&-1&0\\ \end{array}\right),\nonumber \\
&y_{A_4}=\frac{1}{2}\left(\begin{array}{cccccccc} 0&-1&1&-1 &0&0&1&0 \\0&1&1&1 &0&0&1&0\\0&1&1&-1 &0&0&-1&0\\-1&0&0&0 &1&1&0&-1\\
-1&0&0&0 &1&-1&0&1 \\-1&0&0&0 &-1&1&0&1\\-1&0&0&0 &-1&-1&0&-1\\0&-1&1&1 &0&0&-1&0\\ \end{array}\right).
\label{groupA4}
\end{eqnarray}
\normalsize

The relationship between the finite group $A_4$ and the Lie algebra $sl(3,\mathbb{C})$ is established in Sec. \ref{sl3C}.

\section{The Lie algebra of $sl(3,\mathbb{C})$: old and new}
\label{sl3C}

Group operations we considered in our earlier papers were finite group
operations. We are now interested in group operations which are smooth, yet
still compatible with the finite symmetries. This is where the concept of a Lie
group, endowed with its Lie algebra of commutation relations, enters the game.
For quantum mechanics, the favourite Lie group is the matrix Lie group
$SL(n,\mathbb{C})$. For an introduction to Lie groups and Lie algebras see
\cite{Hall03}--\cite{Frappat00}, and the appendix of this paper.

\subsection*{Standard representation of $sl(3,\mathbb{C})$ }

It is well know that $sl(3,\mathbb{C})$ occurs in the context of particle
physics for representing quark states. It is part of the standard model of
elementary particles $su(3) \oplus su(2) \oplus u(1)$ \cite{Carruthers1966}.
Remarkably, one arrives at a form reminiscent of the standard model in
representing the Lie algebra attached to groups $A_4$ and $S_4$, as given in
Sec.\,(\ref{Entangl3}).

A Chevalley basis for the algebra $sl(3,\mathbb{C})$ may be written as
\footnotesize
$$x_1=\left(\begin{array}{ccc} 0 & 0 & 0 \\0 & 0 & 1 \\ 0 & 0 & 0 \\ \end{array}\right),~x_2=\left(\begin{array}{ccc} 0 & 1 & 0 \\0 & 0 & 0 \\ 0 & 0 & 0 \\ \end{array}\right),~x_3=\left(\begin{array}{ccc} 0 & 0 & 1 \\0 & 0 & 0 \\ 0 & 0 & 0 \\  \end{array}\right),$$
\normalsize
\footnotesize
$$y_1=\left(\begin{array}{ccc} 0 & 0 & 0 \\0 & 0 & 0 \\ 0 & 1 & 0 \\ \end{array}\right),~y_2=\left(\begin{array}{ccc} 0 & 0 & 0 \\1 & 0 & 0 \\ 0 & 0 & 0 \\ \end{array}\right),~y_3=\left(\begin{array}{ccc} 0 & 0 & 0 \\0 & 0 & 0 \\ 1 & 0 & 0 \\ \end{array}\right),$$
\normalsize
\small
$$h_1=\left(\begin{array}{ccc} 0 & 0 & 0 \\0 & 1 & 0 \\ 0 & 0 & -1 \\ \end{array}\right),~h_2=\left(\begin{array}{ccc} 1 & 0 & 0 \\0 & -1 & 0 \\ 0 & 0 & 0 \\ \end{array}\right),$$
%
\normalsize and the corresponding table of commutators reads
\begin{equation}
\left(\begin{array}{c|ccc|ccc|cc} [.,.]&x_1&x_2&x_3&y_1&y_2 &y_3&h_1&h_2\\\hline x_1&.&-x_3&.&h_1 &.&y_2&-2x_1&x_1
 \\x_2& &.&.&. &h_2&-y_1&x_2&-2x_2 \\x_3& & &.&x_2 &-x_1&h_1+h_2&-x_3&-x_3\\ \hline y_1& & & &. &y_3&.&2y_1&-y_1\\
y_2& & & & &. &.&-y_2&2y_2 \\y_2& & & & &&.&y_3&y_3\\\hline h_1& & & &  & & &.&.\\h_2& & & &  & & & &.\\ \end{array}\right).
\label{commutator}
\end{equation}
Using this table, the positive roots relative to the pair of generators
$H=(h_1,h_2)$ are easily discerned as $\alpha_1=(2,-1)$, $\alpha_2=(-1,2)$ and
$\alpha_3=(1,1)$, corresponding to the root vectors $x_1$, $x_2$ and $x_3$,
respectively (see \cite{Hall03} for details \footnote{For instance, since
$[h_1,x_1]=2x_1$ and $[h_2,x_1]=-x_1$, one gets the first root
$\alpha_1=(2,-1)$ corresponding to the root vector $x_1$.}). Negative roots
have opposite signs. The Killing matrix is of the form \small
\begin{equation}
\mbox{Kil}=6\left(\begin{array}{cccccccc} 2&.&.&. &1&.&.&. \\.&.&.&1 &.&.&.&. \\.&.&.&. &.&.&1&.\\.&1&.&. &.&.&.&.\\
1&.&.&.&2&.&.&. \\.&.&.&. &.&.&.&1\\.&.&1&. &.&.&.&.\\.&.&.&. &.&1&.&.\\ \end{array}\right).
\end{equation}
\label{KillingMatrix}
\normalsize

The adjoint representation provides another representation of the Lie algebra
$sl(3,\mathbb{C})$
\footnotesize
$$ad_{x_1}=\left(\begin{array}{cccccccc} .&.&.&. &.&.&-2&1 \\.&.&.& .&.&.&.&. \\.&-1&.&. &.&.&.&.\\.&.&.&. &.&.&.&.\\
.&.&.&. &.&1&.&. \\.&.&.&. &.&.&.&.\\.&.&.&1 &.&.&.&.\\.&.&.&. &.&.&.&.\\ \end{array}\right),~~
ad_{x_2}=\left(\begin{array}{cccccccc} .&.&.&. &.&.&.&. \\.&.&.& .&.&.&1&-2 \\1&.&.&. &.&.&.&.\\.&.&.&. &.&-1&.&.\\
.&.&.&. &.&.&.&. \\.&.&.&. &.&.&.&.\\.&.&.&. &.&.&.&.\\.&.&.&. &1&.&.&.\\ \end{array}\right),$$
$$ad_{x_3}=\left(\begin{array}{cccccccc} .&.&.&. &-1&.&.&. \\.&.&.&. &2&.&.&. \\.&.&.&. &.&.&-1&-1\\.&.&.&. &.&.&.&.\\
.&.&.&. &.&.&.&. \\.&.&.&. &.&.&.&.\\.&.&.&. &.&1&.&.\\.&.&.&. &.&1&.&.\\ \end{array}\right),
ad_{y_1}=\left(\begin{array}{cccccccc} .&.&.&. &.&.&.&. \\.&.&-1&. &.&.&.&. \\.&.&.&. &.&.&.&.\\.&.&.&. &.&.&2&-1\\
.&.&.&. &.&.&.&. \\.&.&.&. &1&.&.&.\\-1&.&.&. &.&.&.&.\\.&.&.&. &.&.&.&.\\ \end{array}\right),$$~~
$$ad_{y_2}=\left(\begin{array}{cccccccc} .&.&1&. &.&.&.&. \\.&.&.& .&.&.&.&. \\.&.&.&. &.&.&.&.\\.&.&.&. &.&.&.&.\\
.&.&.&. &.&.&-1&2 \\.&.&.&-1 &.&.&.&.\\.&.&.&. &.&.&.&.\\.&-1&.&. &.&.&.&.\\ \end{array}\right),
ad_{y_3}=\left(\begin{array}{cccccccc} .&.&.&. &.&.&.&. \\.&.&.&. &.&.&.&. \\.&.&.&. &.&.&.&.\\.&1&.&. &.&.&.&.\\
-1&.&.&. &.&.&.&. \\.&.&.&. &.&.&1&1\\.&.&-1&. &.&.&.&.\\.&-1&.&. &.&.&.&.\\ \end{array}\right),$$
$$ad_{h_1}=\left(\begin{array}{cccccccc} 2&.&.&. &.&.&.&. \\.&-1&.&. &.&.&.&. \\.&.&1&. &.&.&.&.\\.&.&.&-2 &.&.&.&.\\
.&.&.&. &1&.&.&. \\.&.&.&. &.&-1&.&.\\.&.&.&. &.&.&.&.\\.&.&.&. &.&.&.&.\\ \end{array}\right),~~
ad_{h_2}=\left(\begin{array}{cccccccc} -1&.&.&. &.&.&.&.\\.&2&. &.&.&.&. \\.&.&1&. &.&.&.&.\\.&.&.&1 &.&.&.&.\\
.&.&.&. &-2&.&.&. \\.&.&.&. &.&-1&.&.\\.&.&.&. &.&.&.&.\\.&.&.&. &.&.&.&.\\ \end{array}\right).
$$
\normalsize


Using the Cartan subalgebra $(h_1',h_2')$, with $h_1'=\mbox{diag}(1,.,1,-1,.,-1,.,.)$ and $h_2'=\mbox{diag}(.,1,1,.,-1.,-1,.,.)$, the new system of positive roots is computed as $\{(1,0),(0,1),(1,1)\}$.

\subsection*{Representation of $sl(3,\mathbb{C})$ stemming from $A_4$}

Let us now go back to tripartite quantum entanglement and show how the
$B$-states (\ref{balanced}) are related to a new representation of
$sl(3,\mathbb{C})$.

Using Magma, we created a (real) subalgebra of the matrix Lie algebra defined
over the rational field, that is obtained from the generators of the finite
group $A_4$ described in (\ref{groupA4}). The algebra is found to be isomorphic
to the Lie algebra $\mathfrak{g}_{A_4}$ of type $sl(3,\mathbb{C})\oplus u(1)$,
and the derived algebra
$\mathfrak{g}_{A_4}'=[\mathfrak{g}_{A_4},\mathfrak{g}_{A_4}]$ turns out to be
isomorphic to $sl(3,\mathbb{C})$.

A Chevalley basis of the algebra $\mathfrak{g}_{A_4}'$ is as follows

\scriptsize
$$x_1=\left(\begin{array}{cccccccc} .&.&.&. &.&.&.&. \\.&.&.&1 &.&.&1&. \\.&.&.&-1 &.&.&-1&.\\.&.&.&. &.&.&.&.\\
.&.&.&. &.&.&.&. \\.&.&.&. &.&.&.&.\\.&.&.&. &.&.&.&.\\.&.&.&. &.&.&.&.\\ \end{array}\right),~~
x_2=\left(\begin{array}{cccccccc} .&1&-1&. &.&.&.&. \\.&.&.&. &.&.&.&. \\.&.&.& .&.&.&.&.\\.&.&.&. &.&.&.&.\\
.&.&.&. &.&.&.&. \\.&.&.&. &.&.&.&.\\.&.&.&. &.&.&.&.\\.&1&-1&. &.&.&.&.\\ \end{array}\right),~~
x_3=2\left(\begin{array}{cccccccc} .&.&.&1 &.&.&1&. \\.&.&.&. &.&.&.&. \\.&.&.&. &.&.&.&.\\.&.&.&. &.&.&.&.\\
.&.&.&. &.&.&.&. \\.&.&.&. &.&.&.&.\\.&.&.&. &.&.&.&.\\.&.&.&1 &.&.&1&.\\ \end{array}\right),
$$
$$y_1=\frac{1}{4}\left(\begin{array}{cccccccc} .&.&.&. &.&.&.&. \\.&.&.&. &.&.&.&. \\.&.&.&. &.&.&.&.\\.&1&-1&. &.&.&.&.\\
.&.&.&. &.&.&.&. \\.&.&.&. &.&.&.&.\\.&1&-1&. &.&.&.&.\\.&.&.&. &.&.&.&.\\ \end{array}\right),~~
y_2=\frac{1}{4}\left(\begin{array}{cccccccc} .&.&.&. &.&.&.&. \\1&.&.& &.&.&.&1 \\-1&.&.& &.&.&.&-1\\.&.&.&. &.&.&.&.\\
.&.&.&. &.&.&.&. \\.&.&.&. &.&.&.&.\\.&.&.&. &.&.&.&.\\.&.&.&. &.&.&.&.\\ \end{array}\right),~~
y_3=\frac{1}{8}\left(\begin{array}{cccccccc} .&.&.&. &.&.&.&. \\.&.&.& &.&.&.&. \\.&.&.& &.&.&.&.\\1&.&.&. &.&.&.&1\\
.&.&.&. &.&.&.&. \\.&.&.&. &.&.&.&.\\1&.&.&. &.&.&.&1\\.&.&.&. &.&.&.&.\\ \end{array}\right),
$$
%
$$h_1=\frac{1}{2}\left(\begin{array}{cccccccc} .&.&.&. &.&.&.&. \\.&1&-1&. &.&.&.&. \\.&-1&1&. &.&.&.&.\\.&.&.&-1 &.&.&-1&.\\
.&.&.&. &.&.&.&. \\.&.&.&. &.&.&.&.\\.&.&.&-1 &.&.&-1&.\\.&.&.&. &.&.&.&.\\ \end{array}\right),~~
h_2=\frac{1}{2}\left(\begin{array}{cccccccc} 1&.&.&. &.&.&.&1\\.&-1&1 &.&.&.&. \\.&1&-1& &.&.&.&.\\.&.&.&. &.&.&.&.\\
.&.&.&. &.&.&.&. \\.&.&.&. &.&.&.&.\\.&.&.&. &.&.&.&.\\1&.&.&. &.&.&.&1\\ \end{array}\right).
$$
\normalsize and its elements are readily seen to fit into the table of
commutators (\ref{commutator}) of $sl(3,\mathbb{C})$. As a result, the roots
relative to a new pair of generators $(h_1,h_2)$ given above are the $\alpha_i$
given in the preceding subsection. This may be a useful feature of the new
representation, in contrast to the adjoint one, for subsequent applications to
the physics of elementary particles.

Going upstream in the group sequence (\ref{sequ}) one arrives at a three-qubit realization of the symmetric group $S_4$. The group $A_4$, with generators as in (\ref{groupA4}), is the derived subgroup of $S_4$. The corresponding Lie algebra, of dimension $13$, may be decomposed as a direct sum of simple Lie algebra as follows
\begin{equation}
\mathfrak{g}_{S_4}=sl(3,\mathbb{C}) \oplus sl(2,\mathbb{C})  \oplus u(1) \oplus u(1),
\label{basissu11}
\end{equation}
in which the algebra $sl(3,\mathbb{C}) \oplus u(1)$ is embedded.

A basis for the representation of $sl(2,\mathbb{C})$ in (\ref{basissu11}) is as follows
\small
$$\left(\begin{array}{cccccccc} 1&.&-1&. &1&-1&.&. \\.&-1&.&. &-1&1&.&1 \\-1&.&1&. &-1&1&.&.\\.&.&.&. &.&.&.&.\\
1&-1&-1&. &.&.&.&1 \\-1&1&1&. &.&.&.&-1\\.&.&.&. &.&.&.&.\\.&1&.&. &1&-1&.&-1\\ \end{array}\right),~~$$
$$\left(\begin{array}{cccccccc} .&1&.&. &1&-1&.&-1 \\.&-\frac{1}{2}&.& .&-\frac{1}{2}&\frac{1}{2}&.&\frac{1}{2} \\.&-1&.&. &-1&1&.&1\\.&.&.&. &.&.&.&.\\
.&\frac{1}{2}&.&. &\frac{1}{2}&-\frac{1}{2}&.&-\frac{1}{2} \\.&-\frac{1}{2}&.&. &-\frac{1}{2}&\frac{1}{2}&.&\frac{1}{2} \\.&.&.&. &.&.&.&.\\.&\frac{1}{2}&.&. &\frac{1}{2}&-\frac{1}{2}&.&-\frac{1}{2}\\ \end{array}\right),
~~
\left(\begin{array}{cccccccc} .&.&.&. &.&.&.&. \\1&-\frac{1}{2}&-1& .&\frac{1}{2}&-\frac{1}{2}&.&\frac{1}{2} \\.&.&.&. &.&.&.&.\\.&.&.&. &.&.&.&.\\
1&-\frac{1}{2}&-1&. &\frac{1}{2}&-\frac{1}{2}&.&\frac{1}{2}
\\-1&\frac{1}{2}&1&. &-\frac{1}{2}&\frac{1}{2}&.&-\frac{1}{2}\\.&.&.&.
&.&.&.&.\\-1&\frac{1}{2}&1&. &-\frac{1}{2}&\frac{1}{2}&.&-\frac{1}{2}\\
\end{array}\right),$$ \normalsize The Killing matrix of the representation may
be diagonalized \small
$$24\left(\begin{array}{ccc} 4 &1 &1 \\1 & . &2 \\ 1 &2 & . \\ \end{array}\right) :=T D T^{-1}~\mbox{with}~D=96\left(\begin{array}{ccc} 1 & . &. \\. & -1 &0 \\ . & . & 3 \\ \end{array}\right)~ \mbox{and}~ T:=\left(\begin{array}{ccc} 1 & . &. \\-1 & 4 & . \\ 2 & -7 & -1 \\ \end{array}\right).$$
\normalsize
corresponding to the representation $su(1,1)$ of $sl(2,\mathbb{C})$, with signature $(2,1)$.

\section{Conclusion}

We have found a new intricate relation between finite group theory, Lie
algebras and three-qubit quantum entanglement. In particular, the connection
between {\it balanced} tripartite entanglement (in Sec. 2) and the
eight-dimensional representation of the Lie algebra $sl(3,\mathbb{C})$ (in Sec.
5) is put forward. Earlier papers of one of the authors focused on the
three-qubit realization of Coxeter groups, such as the largest one $W(E_8)$,
together with its most relevant subgroups comprising the three-qubit Clifford
group $\mathcal{C}_3^+$, $W(E_7)$, $W(E_6)$, $W(F_4)$ and other subgroups
\cite{Planat09,PlanatCPT09}. Here, one discovers that the two-qubit real
entangling gate $S_2$ [see Eq. (\ref{gateS2})] and its three-qubit parent, the
gate $S_3$ [see eq. (\ref{gateS3})] are building stones of the realization of
$W'(E_8)$. An appropriate reduction of $W'(E_8)$ to the four-letter alternating
group $A_4$ [see (\ref{sequ})] is used to represent the algebra
$\mathfrak{g}_{A_4}=sl(3,\mathbb{C}) \oplus u(1)$.
The parent of $A_4$ is the symmetric group $S_4$ and the corresponding Lie
algebra is $\mathfrak{g}_{S_4}=sl(3,\mathbb{C}) \oplus sl(2,\mathbb{C})\oplus
u(1) \oplus u(1)$, which reminds us of the standard model of particles
\cite{Ma87}. From a mathematical point of view, the relation to algebraic
surfaces is worthwhile to be investigated in the future \cite{Yang04}. As an
interesting implication for biosciences, the four letters occurring in the
permutation groups $A_4$ and $S_4$ suggest to consider $\mathfrak{g}_{S_4}$
algebra as a new candidate for a deeper insight into the degeneracies of
genetic code.

\section*{Acknowledgements}

Part of this work was carried out within the framework of the Cooperation Group
``Finite Projective Ring Geometries: An Intriguing Emerging Link Between
Quantum Information Theory, Black-Hole Physics and Chemistry of Coupling" at
the Center for Interdisciplinary Research (ZiF), University of Bielefeld,
Germany. The authors also greatly acknowledge Maurice Kibler for his feedback
and his careful reading of the paper. M. S. was also supported by the VEGA
grants Nos. 2/7012/27 and 2/0092/09.

\section*{Appendix: Elements on Lie groups and Lie algebras}

Let $G$ be a matrix Lie group, the Lie algebra $\mathfrak{g}$ of $G$ is real
and defined as the set of all matrices $X$ such that $e^{tX}$ is in $G$ for all
real numbers $t$. There is an important property that
$$\mbox{for}~\mbox{any}~ X \in \mathfrak{g},~ \mbox{and}~\mbox{for}~ A \in G, ~ \mbox{Ad}_A(X)=AXA^{-1} \in \mathfrak{g},$$
i.\,e. conjugation of an element of the Lie algebra by an element of the Lie
group preserves the algebra. The above map from the Lie algebra to itself is
called the adjoint mapping.

This definition is reminiscent of the definition of the Clifford group
$\mathcal{C}$, that is defined as the normalizer of the Pauli group
$\mathcal{P}$ within the unitary group~$U(n)$, i.e. denoting $X$ an arbitrary
error arising from the Pauli group, and $A$ an element of the Clifford group
\cite{Planat09,PlanatJorrand08,PlanatSole08}
$$\mbox{then for}~\mbox{any}~ X \in \mathcal{P},~ \mbox{and}~\mbox{for}~ A \in \mathcal{C}\subset U(n),~ AXA^{-1} \in \mathcal{P}.$$
In some sense, Lie groups and algebras are a smooth (continuous) formulation of quantum error correction.

The Lie algebra is endowed with a map (called commutator) $[.,.]$ from
$\mathfrak{g} \times \mathfrak{g}$ to $\mathfrak{g}$, with the properties
$$(i)~[.,.]~\mbox{is}~\mbox{bilinear}$$
$$(ii)~[X,Y]=-[Y,X]~\mbox{for}~\mbox{all}~X,Y \in \mathfrak{g}~\mbox{(anticommutativity)}$$
$$(iii)~[X,[Y,Z]]+[Y,[Z,X]]+[Z,[X,Y]]=0~\mbox{for}~\mbox{all}~X,Y,Z \in \mathfrak{g}~\mbox{(Jacobi}~\mbox{identity)}.$$
The adjoint endomorphism \lq\lq Ad" can be reformulated in terms of commutators by the linear map \lq\lq ad" as follows
$$\mbox{ad}_X:\mathfrak{g}\rightarrow \mathfrak{g}~\mbox{defined}~\mbox{by}~ \mbox{ad}_X(Y)=[X,Y].$$
Thus, the map \lq\lq ad" from $X$ to $\mbox{ad}_X$ is a linear map from
$\mathfrak{g}$ to the space $gl(\mathfrak{g})$ of linear operators from
$\mathfrak{g}$ to $\mathfrak{g}$, and there exists a Lie algebra homomorphism
$\mathfrak{g}$ to $gl(\mathfrak{g})$ by the relation
$$\mbox{ad}_{[X,Y]}=[\mbox{ad}_X,\mbox{ad}_Y].$$
Selecting a basis $X_1,\ldots,X_n$ of the $n$-dimensional Lie algebra, for each $i$ and $j$ one obtains
$$[X_i,X_j]=c_{ij}^kX_k,$$
in which the {\it structure constants} $c_{ij}^k$ (with respect to the basis)
define the bracket operation on $\mathfrak{g}$.

For a simple real or complex Lie algebra (see below) there exists a basis,
called the Chevalley basis, for which the structure constants are relative
integers.

A complex Lie algebra $\mathfrak{g}$ is called indecomposable if the only
ideals in $\mathfrak{g}$ are $\mathfrak{g}$ and $\{0\}$, it is called simple if
it is indecomposable and $\mbox{dim}(\mathfrak{g})\ge 2$. The algebra
$\mathfrak{g}$ is called reductive if it is isomorphic to a direct sum of
indecomposable Lie algebras; it is called semi-simple if it is isomorphic to a
direct sum of simple Lie algebras.

A subalgebra of a Lie algebra $\mathfrak{g}$ is a subspace $\mathfrak{h}$ of
$\mathfrak{g}$ such that $[H_1,H_2]\in \mathfrak{h}$ for all $H_1$ and $H_2 \in
\mathfrak{h}$.

If $\mathfrak{g}$ and $\mathfrak{h}$ are Lie algebras, then a linear map
$\phi:\mathfrak{g}\rightarrow \mathfrak{h}$ is called a {\it Lie algebra
homorphism} if $\phi([X,Y])=[\phi(x),\phi(y)]$, for all $X,Y \in \mathfrak{g}$.
If, in addition, $\phi$ is one-to-one and onto, then $\phi$ is called a {\it
Lie algebra isomorphism}.

The adjoint map \lq\lq ad" may be used to figure out the geometry of the Lie
algebra. The composition of two \lq\lq ad" defines a symmetric bilinear form
called the the Killing form

$$B(X,Z)= \mbox{trace} (\mbox{ad}(X)\mbox{ad}(Z)),$$
that possesses several important properties. It is associative, i.\,e.
$B([X,Y],Z)=B(X,[Y,Z])$; it is invariant under the automorphisms $s$ of the
algebra $\mathfrak{g}$ that is, $B(s(X),s(Z))=B(X,Z)$ for $s$ in
$\mbox{Aut}(\mathfrak{g})$, and the Cartan criterion states that a Lie algebra
over a field of characteristic zero is semi-simple iff the Killing form is
nondegenerate.

The matrix elements $B_{ij}$ of the Killing form are related to structure constants as follows

$$B_{ij}=\frac{1}{I_{\mbox{ad}}}c_{im}^n c_{jn}^m,$$
where the {\it Dynkin index} $I_{\mbox{ad}}$ depends on the representation.

\subsection*{Real forms}
A real form is a real Lie algebra $\mathfrak{g}_0$ whose complexification is a
complex Lie algebra $\mathfrak{g}$ \cite{Helgason78}.

Let us define the {\it signature} of a real Lie algebra as a pair $(a_1,a_2)$,
that counts the number of positive ($a_1$) and negative ($a_2$) entries in the
diagonal form of $B$. In particular, a real Lie algebra $\mathfrak{g}$ is
called compact if its Killing form is negative definite. It is also known that
a compact Lie algebra corresponds to a compact Lie group.

As an illustrative example, the special linear algebra $sl(2,\mathbb{C})$ has
two real forms, the so-called (non-compact) split real form
$sl(2,\mathbb{R})\cong su(1,1)$ of signature $(2,1)$ and the compact real form
$su(2)$ of signature $(0,3)$ \footnote{The non-compact split real form $e_7(7)$
of the Lie algebra $e_7$ plays a role in \cite{Levay08}.}. The first real form
$sl(2,\mathbb{R})$ follows from the representation of $sl(2,\mathbb{C})$ in the
Pauli spin basis, with Killing matrix \small $4\left(\begin{array}{ccc} 2 & 0 &
0 \\0 & 0 & 1 \\ 0 & 1 & 0 \\ \end{array}\right)$ \normalsize and eigenvalues
$4$, $8$ and $-4$. The second real form follows from the adjoint representation
[associated to the orthogonal Lie group $SO(3)$] \small
$$\mbox{ad}_{\sigma_z}=\left(\begin{array}{ccc} 0 & -i & 0 \\i & 0 & 0 \\ 0 & 0 & 0 \\ \end{array}\right),~~\mbox{ad}_{\sigma_x}=\left(\begin{array}{ccc} 0 & 0 & 0 \\0 & 0 & -i \\ 0 & i & 0 \\ \end{array}\right),~~\mbox{ad}_{\sigma_y}=\left(\begin{array}{ccc} 0 & 0 & i \\0 & 0 & 0 \\-i & 0 & 0 \\ \end{array}\right),$$
\normalsize
with Killing matrix \small $2\left(\begin{array}{ccc} 1 & 0 & 0 \\0 & 1 & 0 \\ 0 & 0 & 1 \\ \end{array}\right).$\normalsize

More generally, the algebra $sl(3,\mathbb{C})$ has three real forms, the split
real form $sl(3,\mathbb{R})$, the compact real form $su(3)$, and the non-split
real form $su(2,1)$.

\subsection*{Roots}

Let $\mathfrak{g}$ be a complex semi-simple Lie algebra, then a Cartan
subalgebra of $\mathfrak{g}$ is a complex subspace $\mathfrak{h}$ of
$\mathfrak{g}$ with the following properties
$$(i)~\mbox{For}~\mbox{all}~H_1~\mbox{and}~H_2 \in \mathfrak{h}, [H_1,H_2]=0,$$
$$(ii)~\mbox{For}~\mbox{all}~X \in \mathfrak{g},~ \mbox{if}~[H,X]=0~ \mbox{for}~\mbox{all}~H\in \mathfrak{h}, ~\mbox{then}~X \in \mathfrak{h},$$
$$(iii)~\mbox{For}~\mbox{all}~H \in \mathfrak{h},~ \mbox{ad}_H ~\mbox{is}~\mbox{diagonalizable}$$
A root of $\mathfrak{g}$, relative to a Cartan subalgebra of $\mathfrak{h}$, is
a nonzero linear functional $\alpha$ on $\mathfrak{h}$ such that there exists a
nonzero element $X$ of $\mathfrak{g}$ with
$$[H,X]=\alpha(H) X,$$
for all $H$ in $\mathfrak{h}$.

If $\alpha$ is a root, then the {\it root space} $\mathfrak{g}_{\alpha}$ is the
space of all $X$ in $\mathfrak{g}$ for which $[H,X]=\alpha(H)X$ for all $H$ in
$\mathfrak{h}$. The elements of $\mathfrak{g}_{\alpha}$ are the root vectors
for the root $\alpha$. It is convenient to single out the set
$\{\alpha_1,\ldots,\alpha_k\}$ of roots, that have the property that all the
roots can be expressed as linear combinations of the $\alpha_i (i=1..k)$. Such
roots are called {\it positive simple roots}.

As it is well known, the geometry of a semi-simple Lie algebra $\mathfrak{g}$
may be made explicit by introducing the {\it Cartan subalgebra} and its
attached root space. The connection of Lie algebras to finite symmetries occurs
by looking at the isometry group of the root system. Specifically, the subgroup
generated by reflections through the hyperplanes orthogonal to the roots is
called the Weyl group $W(\mathfrak{g})$ [also denoted $W(G)$] of the Lie
algebra $\mathfrak{g}$ (and of the corresponding Lie group $G$).

Weyl groups are found in the context of quantum error correction
\cite{PlanatKibler}. Weyl group $W(E_8)$ and related Weyl subgroups were
already encountered in Sec. (\ref{Entangl3}), in a three-qubit realization.

\end{document}